\documentclass[amsmath,twocolumn,prb,aps]{revtex4-1}
\usepackage{amsthm,amsfonts,graphicx,verbatim, color}
\usepackage{bm}
\usepackage[utf8]{inputenc}

\newcommand{\rot}[1]{\ensuremath{C_{#1}}}
\newcommand{\disp}{\mathcal{E}}
\newcommand{\eq}[1]{Eq.~(\ref{#1})}

\begin{document}

\title{Composite fermions in bands with N-fold rotational symmetry}

\author{Matteo Ippoliti, Scott D. Geraedts, and R. N. Bhatt}

\affiliation{Departments of Electrical Engineering and Physics, Princeton University, Princeton NJ 08544, USA}

\begin{abstract}
We study the effect of band anisotropy with discrete rotational symmetry $\rot{N}$ (where $N\ge 2$) in the quantum Hall regime of two-dimensional electron systems.
We focus on the composite Fermi liquid (CFL) at half filling of the lowest Landau level.
We find that the magnitude of anisotropy transferred to the composite fermions decreases very rapidly with $N$.
We demonstrate this by performing density matrix normalization group calculations on the CFL, and comparing the anisotropy of the composite fermion Fermi contour with that of the (non-interacting) electron Fermi contour at zero magnetic field.
We also show that the effective interaction between the electrons after projecting into a single Landau level is much less anisotropic than the band, a fact which does not depend on filling and thus has implications for other quantum Hall states as well. 
Our results confirm experimental observations on anisotropic bands with warped Fermi contours, where 
the only detectable effect on the composite Fermi contour is an elliptical distortion ($N = 2$).
\end{abstract}

\maketitle

\section{Introduction}

Two-dimensional electron gases in strong perpendicular magnetic fields exhibit a range of interesting behavior which has engaged condensed matter physics for the past several decades\cite{JainBook,PrangeGirvin}.
These phenomena go under the name of ``quantum Hall effect'', an umbrella term which includes gapped states (both Abelian and non-Abelian) as well as the gapless composite Fermi liquid, found when the lowest Landau level is half-filled (filling factor $\nu = 1/2$). 
One topic of recent interest is understanding such gapless states in the absence of rotational symmetry. 
Though much of the research on quantum Hall systems has made the assumption of rotational symmetry, this is not essential to the quantum Hall effect\cite{Haldane2011,Can2014,Bradlyn2015,Gromov2015}; 
consequently, such an understanding should be possible.
Rotational symmetry may be broken theoretically by introducing a metric into the Hamiltonian through either the dispersion relation or dielectric tensor. 
This approach yields an elliptical anisotropy, which has been studied previously by several groups\cite{Qiu2012,BoYang2012,Wang2012,Apalkov2014}.
In this work we refer to this as a `2-fold' anisotropy, since ellipses are invariant under the group of $\pi$ rotations $\rot{2}$.

The effects of anisotropy can also be probed experimentally, e.g. by straining samples\cite{Jo2017} or studying electron systems in semiconductors which have anisotropic dispersion relations
\cite{Shayegan2006,Gokmen2010}. 
In high magnetic field at half-filling of the lowest Landau level, such experiments realize a `composite Fermi liquid'\cite{JainBook,HLR}: 
a compressible phase where composite fermions, consisting of an electron and two flux quanta, see no effective magnetic field at the mean-field level, and form a `composite Fermi sea'. 
The outline of this composite Fermi sea (a composite Fermi contour) can be detected experimentally and compared to the zero-field electron Fermi contour. 

The dispersion relations of carriers in experiment lead to zero field Fermi contours that are more complicated than simple ellipses. 
Nonetheless, the composite Fermi contours observed experimentally do not seem to exhibit measurable deviations from an ellipse
\footnote{Since experiments can only measure the dimensions of the Fermi sea in two perpendicular directions, making definitive statements on the shape of the composite Fermi sea is difficult. But the assumption that the measured dimensions represent the major and minor axes of an ellipse leads to a Fermi sea with the correct area\cite{Jo2017}.}.
Furthermore, the shape of the composite Fermi contour can also be determined numerically\cite{Geraedts}, and numerical studies of a composite Fermi liquid with elliptical anisotropy are found to agree with the experimental observations\cite{Ippoliti1}. 

One way to discuss the response of a system to anisotropy is in the language of the `intrinsic metric'\cite{Haldane2011}. 
The idea is that a quantum Hall state contains a metric, distinct from that introduced by curved space or that from an anisotropic Hamiltonian (which could be due to an anisotropic dispersion relation or dielectric tensor). 
The intrinsic metric is dynamical, and measuring it is a way to quantify the system's response to externally imposed anisotropy. 
For elliptical anisotropies the intrinsic metric appears as a variational parameter in model wavefunctions\cite{Qiu2012}. 
By finding the value of this parameter which maximizes the overlap between exact states and model states, one can determine the dependence of the intrinsic metric on externally applied anisotropy for gapped states\cite{BoYang2012}. 
For composite Fermi liquid states, the intrinsic metric gives the shape of the composite Fermi contour\cite{Balram2016}, and therefore using the techniques described above, it can be measured both numerically and experimentally.

Another language with which to describe anisotropy is that of the `generalized pseudopotentials'\cite{BoYang2017}. 
The Haldane pseudopotentials\cite{HaldanePP} are a decomposition of any interaction into its action on pairs of electrons with specified relative angular momentum, and therefore their definition requires rotational symmetry. 
Generalized pseudopotentials provide a way to achieve a similar decomposition in the absence of rotational symmetry. 

Our goal in this work is to better understand the response of a quantum Hall system to the presence of band anisotropy beyond the simplest elliptical case. 
The anisotropies we will consider have a discrete $N$-fold rotational symmetry (where $N$ is an even integer), i.e. they are invariant under $\rot{N}$.
This is a generalization of the elliptical case (obtained for $N = 2$), which has been the focus of most of the previous work on anisotropy in quantum Hall states. 
One exception, in which bands with $\rot{4}$ symmetry are considered, is Ref.~\onlinecite{Haldane2016}.

There are generally two ways to introduce anisotropy in a quantum Hall system: 
either via the one-electron dispersion, or via the electron-electron interaction.
In the case of elliptical ($N=2$) anisotropy, the two approaches are equivalent, since a linear coordinate transformation can move the anisotropy between the band mass tensor and the dielectric tensor. 
For more complicated anisotropies no linear transformation can accomplish this, and 
thus band anisotropy and interaction anisotropy are not identical.

In this work we concentrate on the former, for reasons that will become clear later (though we also consider anisotropic interactions in Section~\ref{sec:results}).
We consider a simple set of dispersions, one for each \rot{N} symmetry group.
Completely general anisotropies such as those studied in experiments can then be thought of as arising from a sum of terms with different $\rot{N}$ symmetries,
so the response of the CFs can be broken down into their response to each individual ``harmonic''.
 
We quantify the anisotropy by $\alpha_F$ (the `anisotropy parameter'), which is the ratio of the largest and smallest Fermi momenta for the system at zero magnetic field (we choose dispersions such that $\alpha_F$ is independent of the Fermi energy). 
Elliptical anisotropy of a given $\alpha_F$ can be obtained by introducing the unimodular metric $diag[ \alpha_F, 1/\alpha_F]$ into the electron mass. 
We can then define $\alpha_{CF}$, which parameterizes the quantum Hall system's response to applied anisotropy, by writing the intrinsic metric as $diag[ \alpha_{CF}, 1/\alpha_{CF}]$. 
For the composite Fermi liquid case, $\alpha_{CF}$ is precisely the ratio of the largest and smallest Fermi momenta of the {\it composite} Fermi surface. 
Going beyond the elliptical case, we can no longer use the language of metrics, but for CFL states we can still define $\alpha_{CF}$ as the ratio of Fermi momenta. 
At present it is not known how to generalize Ref.~\onlinecite{Qiu2012} by writing a model wavefunction which shows more general anisotropy, or how to generalize the coordinate transformation of Ref.~\onlinecite{BoYang2017}.
Therefore comparing the ratio of Fermi momenta in gapless states is the only available method to quantitively study the effects of more general anisotropy.

In Section \ref{sec:model} we formulate dispersion relations with $N$-fold anisotropy and determine the corresponding form factors. 
We express the effective interaction between one-electron orbitals with these form factors in terms of generalized pseudopotentials. We find that for fixed $\alpha_F$ the strength of the anisotropic pseudopotentials decreases rapidly with $N$. 
In Section \ref{sec:results} we present the results of a numerical density matrix renormalization group (DMRG) study with anisotropic dispersion, in particular we show $\alpha_{CF}$ vs $\alpha_F$ for $N=2, 4$, and $6$. 
In agreement with the result for anisotropic pseudopotentials, we find that the dependence of $\alpha_{CF}$ on $\alpha_F$ is sharply reduced as $N$ is increased. 
Our results suggest that anisotropies with $N>2$ have a very small effect on quantum Hall systems, for reasonable values of $\alpha_F$ such as those encountered in experiment.

\section{Model}
\label{sec:model}

We study a system of electrons in two dimensions in the presence of a perpendicular magnetic field with the following Hamiltonian:
\begin{equation}
H = \sum_i \disp(\vec \Pi_i) + H_{\rm int},
\label{eq:ham}
\end{equation}
where $\vec\Pi\equiv \vec p-e\vec A/c$ is the kinetic momentum for electrons in a magnetic field, $\disp(\vec \Pi)$ is the electron dispersion and $H_{\rm int}$ is the interaction between electrons. 
Unless explicitly indicated otherwise, for the rest of this work we  take $H_{\rm int}$ to be the Coulomb interaction.
Further, we study the simplest dispersions that have \rot{N} symmetry, which are homogeneous polynomials\footnote{Since the components of $\vec\Pi$ do not commute with each other, all mixed terms of the form $\Pi_x^m \Pi_y^{N-m}$ must be symmetrized.} of degree $N$ in the kinetic momenta $\Pi_x, \Pi_y$.
The family of such polynomials that are bounded from below (giving rise to a stable dispersion) can easily be constructed as follows.

Consider the function 
\begin{equation}
\disp_{\text{ani}}(\vec k) \equiv k^N \cos(N\theta).
\end{equation} 
where $(k,\theta)$ are polar coordinates in two-dimensional momentum space.
This clearly has the desired \rot{N} symmetry.
Moreover, by observing that $k^N \cos(N\theta) = \text{Re} (k e^{i \theta})^N = \text{Re}(k_x + i k_y)^N$,
we see that this is a homogeneous polynomial of degree $N$ in $k_x$, $k_y$.
The family of \rot{N}-symmetric energy dispersions we consider is then given by
\begin{equation}
\disp (\lambda_F; \vec \Pi) 
= \frac{E_0 \ell_0^N}{\hbar^N} \left[ |\vec \Pi |^N + 
\lambda_F \disp_{\text{ani}} (\vec\Pi) \right] \;,
\label{eq:disp}
\end{equation}
where $E_0$ and $\ell_0$ are arbitrary units of energy and length, respectively.
Requiring this function to be bounded from below (so that the Hamiltonian \eq{eq:ham} has a well-defined ground state) forces $-1 < \lambda_F < +1$.

The zero field electron Fermi contour for the dispersion in \eq{eq:disp} is given by
\begin{equation}
k_F(\theta) = \frac{A}{\ell_0} \left( \frac{1}{1+\lambda_F \cos(N\theta)} \right)^{1/N}
\label{eq:kF_theta}
\end{equation}
where the dimensionless prefactor in \eq{eq:kF_theta} is $A= (E_F/E_0)^{1/N}$, $E_F$ being the Fermi energy.
The dispersions $\disp(-\lambda_F)$ and $\disp (\lambda_F)$ are related by a $\pi/N$ rotation, so we can assume $\lambda_F \geq 0 $ without loss of generality.

The anisotropy parameter $\alpha_F$ can thus be expressed in terms of $\lambda_F$:
\begin{equation}
\alpha_F = \frac{k_F(\pi/N)}{k_F(0)} =  \left( \frac{1+\lambda_F}{1-\lambda_F} \right)^{1/N} \;. 
\label{eq:alpha_lambda}
\end{equation}

While other anisotropic dispersions are possible, the one we have chosen to study has the advantage of being homogenous, 
which makes the shape of the Fermi surface (and therefore $\alpha_F$) independent of the Fermi energy, as is clear from \eq{eq:kF_theta}, 
since $E_F$ appears only in the prefactor $A$. 
We contend that the central result of this work, namely the rapid decrease with $N$ of the effect of $N$-fold anisotropy of the zero-field Fermi contour on the composite Fermi contour (CFC), is generic, and not particular to our form of anisotropic dispersion.

As in the usual isotropic case, the single-particle solutions to Eq.~(\ref{eq:disp}) are organized into degenerate `Landau levels', 
which we denote with tildes ($\tilde{0}, \tilde{1}, \tilde{2}\dots$) to distinguish them from the isotropic Landau levels ($0, 1, 2\dots$).
Once we project to the lowest anisotropic Landau level $\tilde{0}$, electrons interact via the effective potential\cite{BoYang2017}:
\begin{equation}
V(\vec q)=|F_{\tilde 0}(\vec q)|^2 V^{\rm bare}(\vec q),
\label{eq:pot}
\end{equation}
where $V^{\rm bare}(\vec q)$ is the original interaction in momentum space (e.g. for the Coulomb interaction used in this work $V^{\rm bare}(\vec q)=2\pi/|\vec q|$), 
and $F_{\tilde 0}(\vec q)$ is the `form factor' of the $0^{\rm th}$ anisotropic Landau level which is determined from the single-particle degenerate ground state of Eq.~(\ref{eq:disp}). 
The anisotropic form factors can be expressed in terms of the isotropic ones, which result from using $\disp(k)\propto|k|^2$, as follows.
In our dispersion \eq{eq:disp}, we replace $(\ell_B/\hbar) \Pi_x \mapsto (a+a^\dagger)/\sqrt 2$, $(\ell_B/\hbar) \Pi_y \mapsto (a-a^\dagger)/i \sqrt 2$, in terms of the Landau cyclotron annihilation operator $a$.
Here $\ell_B \equiv \sqrt{\hbar/eB}$ is the magnetic length.
Then, expressed in the basis $\{ |n\rangle : n \in \mathbb N_0 \}$ of isotropic Landau levels, the kinetic energy Hamiltonian becomes a tri-diagonal matrix (with entries in diagonals 0 and $\pm N$), whose low-lying energy states can easily be found with a sparse eigensolver.
This provides the coefficients $\{ u_n \}$ in the basis expansion
$ | \tilde{0} \rangle = \sum_{n = 0}^\infty u_n | n \rangle \;, $
which in turn give the desired form factor:
\begin{equation}
F_{\tilde 0}(\vec q) = \sum_{j_1, j_2} u_{j_1} u_{j_2}^* F_{j_1, j_2} (\vec q) \;.
\label{eq:gen_ff}
\end{equation}
The isotropic form factors $F_{j_1, j_2} (\vec q)$ in the above formula are expressible analytically in terms of generalized Laguerre polynomials\cite{PrangeGirvin}.

\begin{figure}
\centering
\includegraphics[width = 0.95\columnwidth]{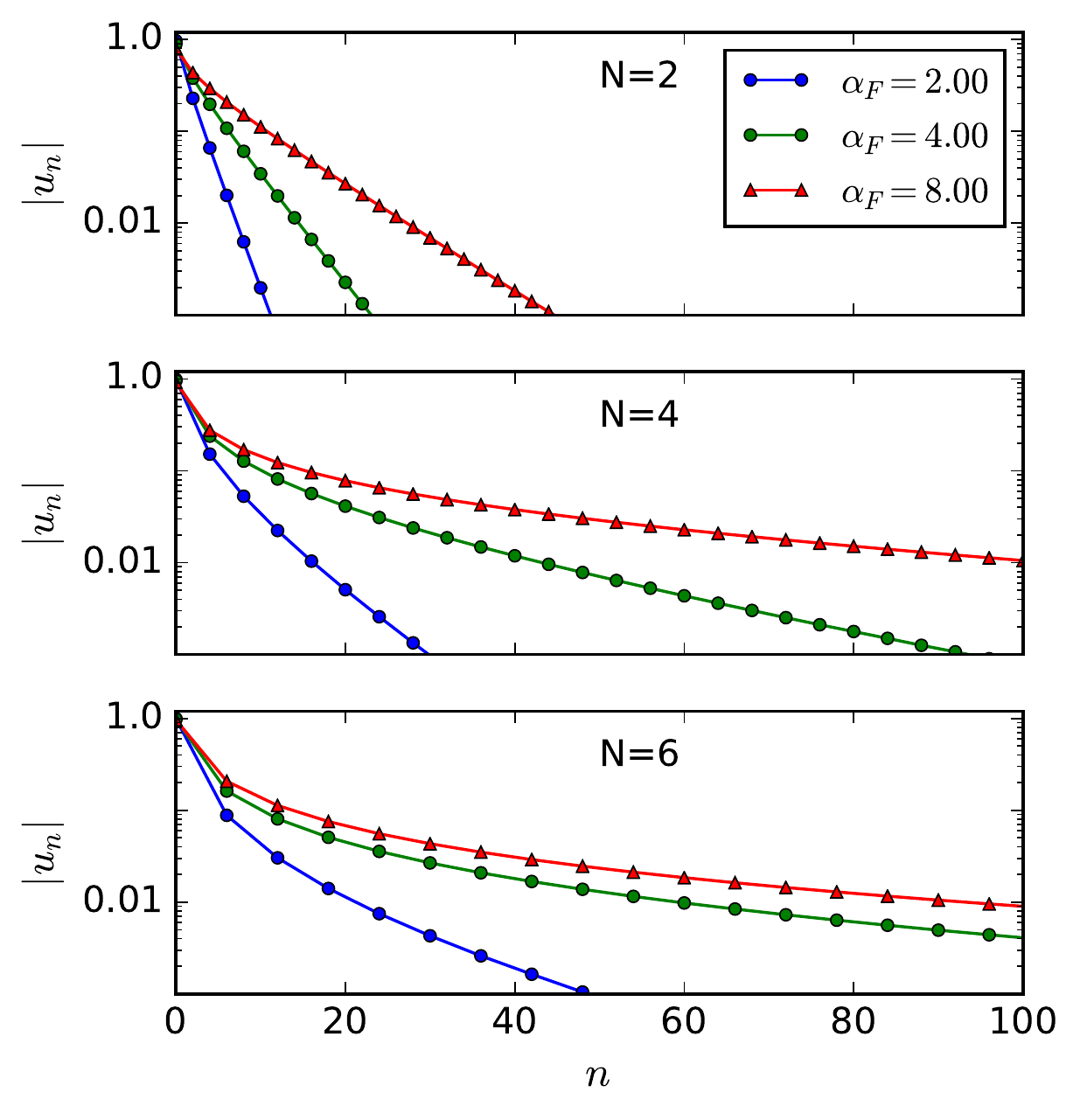}
\caption{Landau level mixing coefficients $u_n$ for the ground state of the two-, four- and six-fold symmetric dispersion in a high magnetic field, 
at several values of the Fermi contour anisotropy $\alpha_F$.
Larger anisotropy leads to a stronger mixing with higher Landau levels. 
Only Landau levels with $n$ which is a multiple of $N$ are involved.}
\label{fig:LLmix}
\end{figure}

An example of the Landau level mixing coefficients $u_n$ as a function of the anisotropy parameter $\alpha_F$ and the rotational symmetry $N$ is given in Figure~\ref{fig:LLmix}.
A few properties of the mixing are apparent from the figure.
First of all, a \rot{N}-symmetric dispersion only mixes Landau levels whose indices differ by an integer multiple of $N$; in particular, the ground state $|\tilde{0}\rangle$ is a mixture of $\{ | 0 \rangle, | N\rangle, | 2N \rangle, \dots\} $.
Secondly, the amplitudes $u_n$ decay exponentially at large $n$, and the decay gets slower for increasing $\alpha_F$ and $N$.
Finally, the leading anisotropic contribution ($n=N$) gets weaker for increasing $N$.

\section{Anisotropic Pseudopotentials}
\label{sec:ani_pp}

It is instructive to express the effective potential \eq{eq:pot} in terms of the generalized pseudopotentials of Ref.~\onlinecite{BoYang2017}. 
Such pseudopotentials can be written as $V_{m,n}^\sigma$, with $m,n$ non-negative integers and $\sigma=\pm$. The generalized pseudopotentials with $n=0,\sigma=+$ are the traditional Haldane pseudopotentials, and for isotropic interactions they are the only terms that appear. 
For $N$-fold anisotropy, terms with $n\ge N$ appear in addition to $n=0$. 
As in the case of traditional pseudopotentials, only terms with odd $m$ contribute for fermions.

In this approach, the true anisotropy of the system in a high magnetic field is best estimated not from the dispersion relation in \eq{eq:disp} but from the effective potential in \eq{eq:pot},
whose anisotropy can be quantified by means of the generalized pseudopotentials. 
Since they form a complete basis, we can expand \eq{eq:pot} in terms of these pseudopotentials and compare the size of the contributions at $n=0$ (the isotropic part) with those at $n>0$.  
Some technical comments are in order.
The effective interaction \eq{eq:pot} is not normalizable, i.e. $\int d^2 q (V(\vec q))^2 = \infty$, due to the $q\to 0$ singularity in the Coulomb potential\footnote{In reality, this singularity is removed by the presence of counter ions.}.
Luckily this singularity only appears in the $m=n=0$ pseudopotential, whereas the only terms which are relevant to the physics of our fermionic system are those with odd $m$, $n$ multiple of $N$, and $\sigma = +$ ($\sigma = - $ would appear if we rotated our coordinates).
We thus measure anisotropy as follows: 
we compute the coefficients 
\begin{equation}
C^+_{m,n} \equiv \int d^2 q V(\vec q) V_{m,n}^+(\vec q),
\end{equation}
then we compute the total norm of terms at a given $n$ as 
\begin{equation}
\chi_n \equiv \sqrt{\sum_{m \text{ odd}} (C_{m,n}^+)^2} ,
\end{equation}
and finally we normalize it by the norm of the isotropic part, $\chi_0$, computed in the same way.
The sum over $m$ should range over all odd positive integers, but we cut it off when numerical convergence is achieved  (typically $m \lesssim 100$).

\begin{figure}
\centering
\includegraphics[width = 0.95\columnwidth]{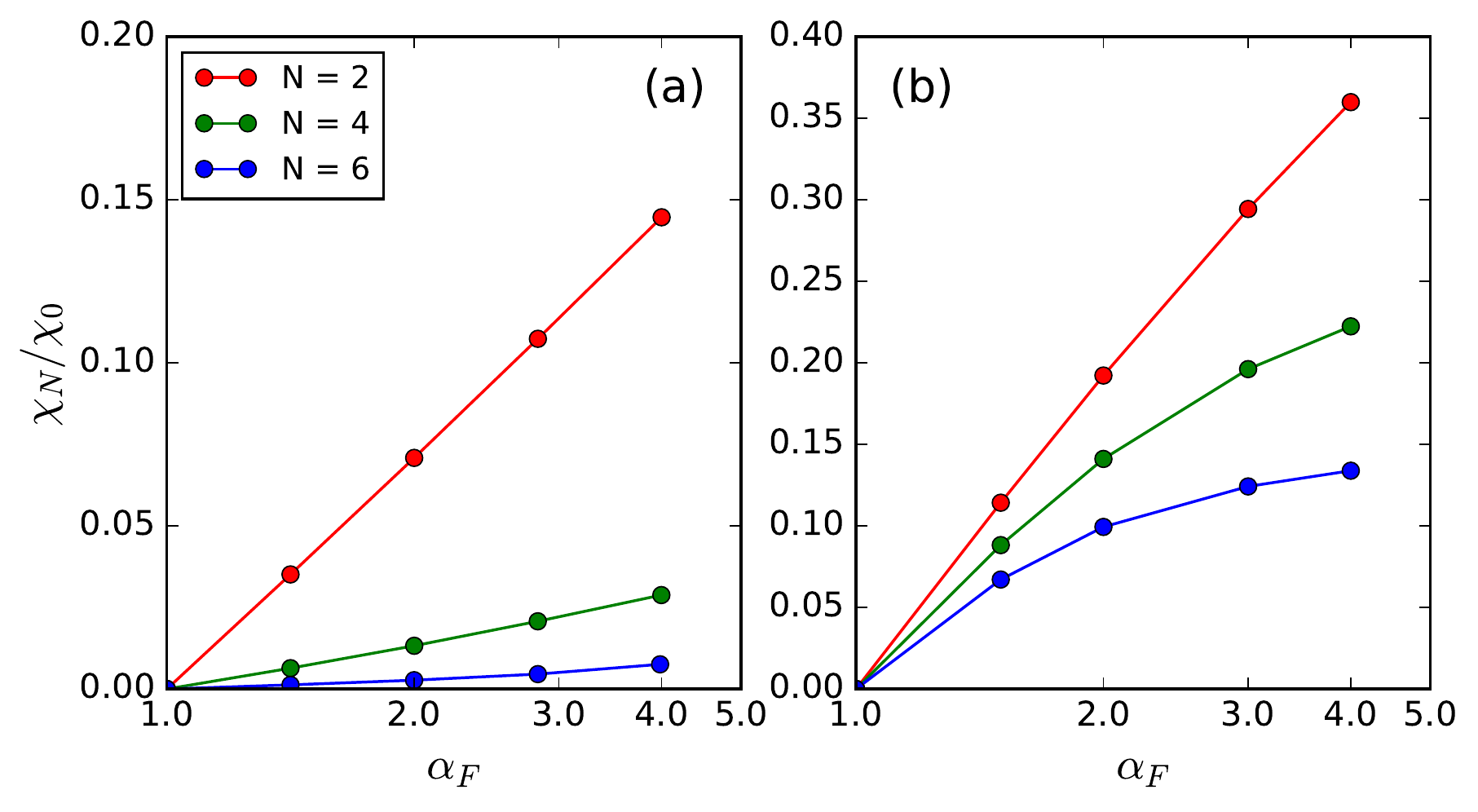}
\caption{(a) Relative weight of leading anisotropic pseudopotentials in the effective interaction \eq{eq:pot}, as a function of anisotropy $\alpha_F$, for $N=2$, 4 and 6. 
The quantity $\chi_N / \chi_0$ is defined in the main text.
(b) Same quantity plotted for the anisotropic Coulomb interaction defined in \eq{eq:anicoul}.
In both cases the growth with $\alpha_F$ is slow (logarithmic or sub-logarithmic) and the effect decreases quickly for increasing $N$.
}
\label{fig:pp}
\end{figure}

Our results for the leading anisotropic contribution, $\chi_N / \chi_0$, are shown in Figure~\ref{fig:pp} and reveal that the effect of a given kinetic energy anisotropy becomes rapidly smaller for increasing $N$, 
i.e. the effective Hamiltonian \eq{eq:pot} is more robust to higher angular momentum distortions.
Importantly, this statement is true at the Hamiltonian level, and thus in principle applies to states at {\it all} fillings. 
A practical consequence is that, for gapped states, anisotropy-driven transitions to states with discrete \rot{N} rotational symmetry ($N > 2$) are expected to occur only at very large values of the symmetry-breaking field
\footnote{This is under the assumption that said transitions occur when the effective interaction has anisotropic pseudopotential components of order $O(1)$.}, 
if at all.
For the gapless CFL state at filling $\nu = \frac12$, the expectation is that the Fermi contour should be significantly stiffer than it is for the elliptical ($N = 2$) case\cite{Ippoliti1}, where $\alpha_{CF} \sim \alpha_F^{0.5}$.
This expectation is tested numerically in the next section.

We can also ask what happens when the interaction, rather than the dispersion relation, is anisotropic. 
We should point out that this scenario is not as well-motivated experimentally. 
The most natural way to engineer an anisotropic interaction with \rot{N}-symmetry in a real system would be to use a cold atomic system composed of `molecules' of positively and negatively charged atoms with $N$-fold symmetry. 
Note that the molecules cannot be neutral, since in that case they could orient themselves to have attractive interactions, and since in a single Landau level there is no kinetic energy they would form bound states instead of a quantum Hall phase. 
For molecules that have net charge, their long ranged interactions will be dominated
by the isotropic Coulomb repulsion (which goes as $r^{-1}$) rather than the anisotropic $N$-fold symmetry part (which comes from the multipolar nature of the molecules and decays as $r^{-(N +1)}$ ). 
Therefore such molecules will have only weakly anisotropic interactions. 
We attempted to make an analog of Fig.~\ref{fig:pp}(a) for such molecular objects,
and found that increasing $N$ by $2$ decreased $\chi_N/\chi_0$ by
approximately four orders of magnitude.

Though it is less physically motivated, we can also
study an anistropric interaction by making the following change to the Coulomb interaction:
\begin{equation}
\frac{1}{r}\rightarrow \frac{1}{r[1+\lambda \cos (N\theta)]^{1/N} }\;.
\label{eq:anicoul}
\end{equation}
This yields equal-potential contours similar to the Fermi
contours described by Eq. (\ref{eq:kF_theta}). $\chi_N/\chi_0$ for such potentials
are shown in Fig.~\ref{fig:pp}(b). We see qualitatively similar behavior as for anisotropic dispersion relations, with $\chi_N/\chi_0$ 
growing sublinearly with $\alpha_F$ and shrinking as $N$ is increased.
Thus we see that the interaction anisotropy, for $N>2$, is small for physical models, or qualitatively similar to anisotropic dispersion for less physically motivated models.
Consequently, in the next Section, we consider only anisotropy arising from the dispersion. 

\section{DMRG Results}
\label{sec:results}

In this section we numerically calculate $\alpha_{CF}$ as a function of $\alpha_F$ for $N = 4$ and 6. 
We obtain the ground state of a system of electrons with dispersion relation given in \eq{eq:disp} and Coulomb interactions, on an infinite cylinder, using density matrix renormalization group (DMRG) for a half-filled lowest Landau level. 
The application of DMRG to quantum Hall systems has been described elsewhere\cite{Zaletel2013};
the only modification we make to this procedure is replacing the standard Landau level form factors with those found in \eq{eq:gen_ff}. 

\begin{figure*}
\centering
\includegraphics[width = 0.9\textwidth]{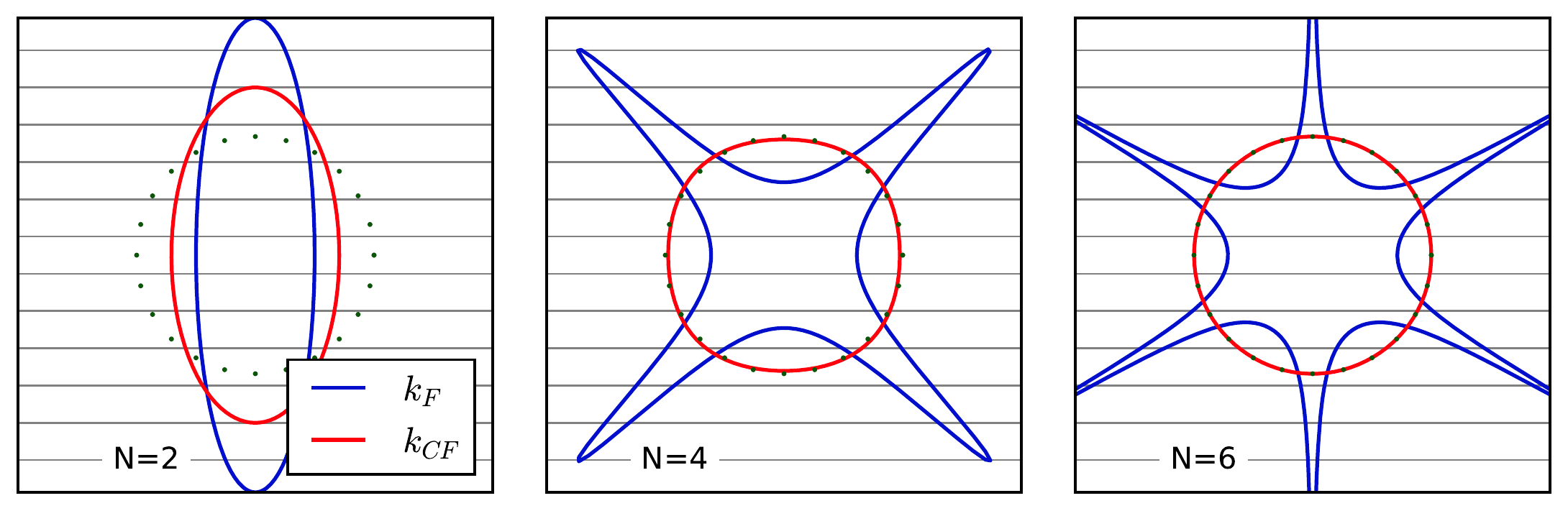}
\caption{Fermi contours for the fermions (blue) and the composite fermions (red), for different values of $N$,
for the same value of zero-field Fermi contour anisotropy, $\alpha_F = 4$.
$\alpha_{CF}$ is obtained from Fig.~\ref{fig:aa}. 
The green dotted circle shows an isotropic Fermi contour for reference. 
The gray lines indicate the accessible momenta on a cylinder of diameter $L_y=20$ magnetic lengths. 
The guiding center structure factor is singular (when $q_y=0$) at values of $q_x$ for which the gray lines connect two parts of the composite fermion Fermi contours, and we use this feature to determine $\alpha_{CF}$.
}
\label{fig:surfaces}
\end{figure*}

Once we have obtained the ground state, we map out its Fermi contour by locating singularities in the guiding center structure factor, a method which has been used successfully on systems with circular and elliptical composite Fermi surfaces\cite{Geraedts,Ippoliti1}. 
To summarize the method, on an infinite cylinder the allowed momenta values are continuous in the $x$ direction but discrete  in the $y$ direction (see Fig.~\ref{fig:surfaces}). 
If we measure the guiding center structure factor $S(q_x,q_y)$ at fixed $q_y=0$, there will be a singularity whenever $q_x$ corresponds to the distance between different points on the Fermi contour along one of the gray lines in Fig. \ref{fig:surfaces}. By finding all such lines we then estimate the shape of the Fermi surface.

\begin{figure*}
\centering
\includegraphics[width = 0.9\textwidth]{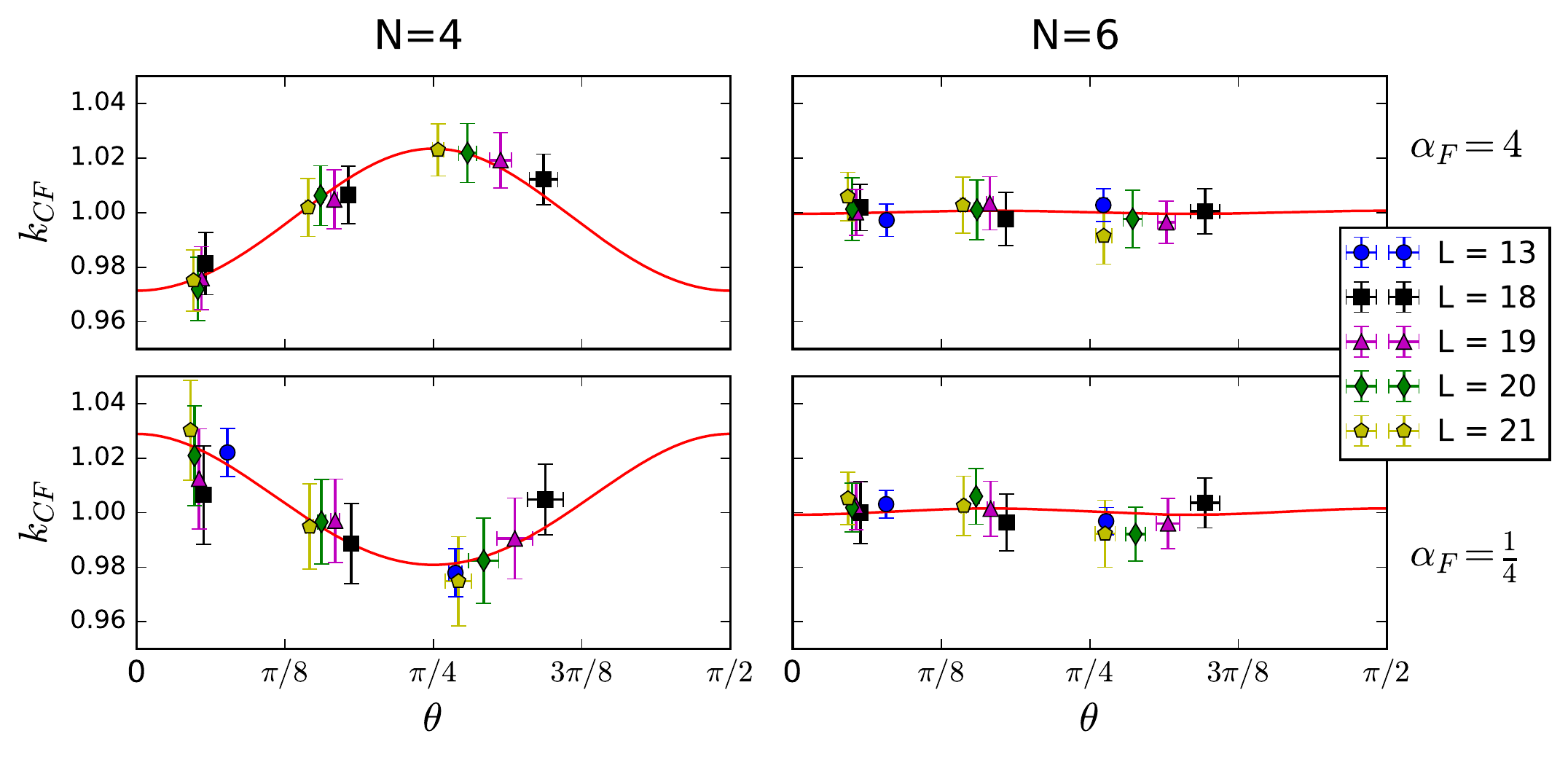}
\caption{Fermi wavevector $k_{CF}$ as a function of angle $\theta$, for $N = 4$ (left) and $N = 6$ (right) and band anisotropy $\alpha_F = 4$ (top) and $\alpha_F = 1/4$ (bottom).
Changing $\alpha_F \mapsto 1/\alpha_F$ rotates the dispersion by $\pi / N$ and thus these values of $\alpha_F$ provide a consistency check for our method.
The solid line represents a fit to $k_{CF}(\theta) = A/(1+\lambda_{CF} \cos(N\theta))^{1/N}$[Eq.~(\ref{eq:kF_theta})].
From $\lambda_{CF}$ one can estimate $\alpha_{CF}$ via \eq{eq:alpha_lambda}.
The $N=4$ data shows anisotropy, whereas the $N = 6$ Fermi contour appears circular within numerical accuracy (i.e. the effect of band anisotropy on the CFs, if present, is too small to be detected). 
}
\label{fig:kF}
\end{figure*}

Examples of the data that we gather with this method are shown in Figure~\ref{fig:kF}.
In the figure we plot the locations of the observed singularities in polar coordinates, i.e. the Fermi momentum $k_F$ as a function of the angle around the Fermi contour $\theta$. 
The resulting data should satisfy Eq.~(\ref{eq:kF_theta}), but with $\lambda$ replaced by $\lambda_{CF}$, from which $\alpha_{CF}$ can be derived using Eq.~(\ref{eq:alpha_lambda}). 
Each plot aggregates singularities from different system sizes
(three to five distinct values of the cylinder circumference ranging between $13\ell_B$ and $21\ell_B$) at the same $N$ and $\alpha_F$.
The prefactor $A$ in Eq.~(\ref{eq:kF_theta}) for the composite fermions is set by Luttinger's theorem. 
In a purely two-dimensional system, the Fermi contour would need to enclose an area of $\pi \ell_B^{-2}$ in momentum space, and this would set $A=1$.
At the finite sizes we can simulate, our requirement is instead that the sum of the length of all the cuts across the Fermi surface totals $L\ell_B^{-2}/2$. 
$A$ is modified from $1$ to satisfy this constraint.
In practice, we find this deviation is typically of order $1$ to $2\%$, which can be comparable in magnitude to the modulation that we seek to measure.
Therefore we rescale each $k_F$ by the average of all the $k_F$'s measured at that size.
This allows a more accurate comparison of different system sizes.

\begin{figure}
\centering
\includegraphics[width = \columnwidth]{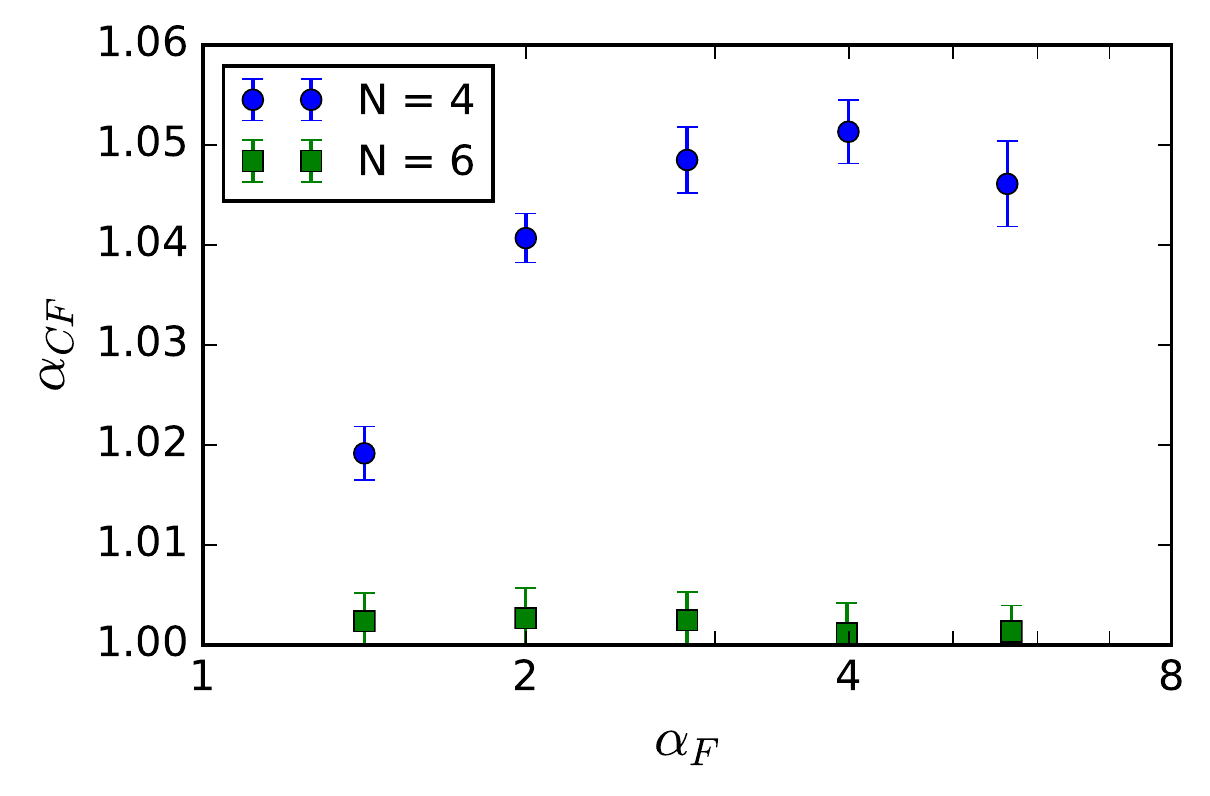}
\caption{CF Fermi contour anisotropy $\alpha_{CF}$, extracted from fits such as those shown in Figure~\ref{fig:kF}, for $\alpha_F$ in the range of 1 to $4\sqrt 2$, for the four- and six-fold symmetric one-electron dispersions defined in \eq{eq:disp}. 
Each datapoint is obtained by combining the estimates obtained at $\alpha_F$ and $1/\alpha_F$, which are related by a $\pi/N$ rotation (as in the top and bottom panels of Figure~\ref{fig:kF}).
}
\label{fig:aa}
\end{figure}

By fitting the data points at each $\alpha_{CF}$ to Eq.~(\ref{eq:kF_theta}),
we can extract the CF Fermi contour anisotropy parameter $\alpha_{CF}$ as a function of $\alpha_F$, the results of which we show in Figure~\ref{fig:aa}.
We see that the dependence of $\alpha_{CF}$ on $\alpha_F$ gets notably weaker as $N$ is increased.
While the results for $N=2$ suggest\cite{Ippoliti1} 
$\alpha_{CF} \approx \alpha_F^{0.5}$, 
in this case we consistently find $\alpha_{CF}$ to be very close to 1 for $N = 4$ and 6.
In particular, the data for $N = 4$ deviates from a power law, showing saturation at $\alpha_{CF} \simeq 1.05$,
while the data for $N = 6$ is entirely consistent with $\alpha_{CF} = 1$ within numerical uncertainty.
This is not surprising based on the drastic reduction in the magnitude of the effect when going from $N = 2$ to $N = 4$.

\section{Discussion}

We have studied a two-dimensional electron gas at half-filling of the lowest Landau level, in the case where rotational symmetry is broken due to an anisotropic dispersion relation for the electrons. 
We investigated how the anisotropy of the zero-field dispersion is transferred to the composite fermions.
In particular we studied anisotropies with $\rot{N}$ symmetry, and found that its effect on the composite Fermi surface decreases rapidly with $N$.
We arrived at this conclusion both by analyzing the effective interaction using generalized pseudopotentials, and by performing numerical studies using DMRG to directly measure the anisotropy of the ground state. 
This result is consistent with our recent finding\cite{Ippoliti2} that CFs are completely insensitive to circularly symmetric band deformations (which can be thought of as the $N \to\infty$ limit of the present set of models), even when the zero-field Fermi contour goes from being a circle to an annulus or a set of disconnected annuli.

Experimental systems with anisotropy would typically require modeling as a sum of such $\rot{N}$-symmetric dispersions.
Our results show that reasonable approximations can be obtained simply by focusing on the smallest value of $N$, namely $N=2$, corresponding to band mass anisotropy (elliptical distortion of the Fermi contour).

In the numerical part of this paper (Sections~\ref{sec:ani_pp} and \ref{sec:results}) we studied a gapless system, namely the composite Fermi liquid at half-filling. 
However our results for the effective interactions should apply to any quantum Hall state.
They suggest that for all quantum Hall states, the response to anisotropy drops rapidly with $N$. 
It would be interesting to test this prediction numerically for $N>2$. 
One barrier to doing this is that the unlike the $N=2$ case, for $N>2$ we do not know how to define the parameter $\alpha_{CF}$ for a gapped quantum Hall state. 
For $N=2$, in addition to being related to the shape of the composite Fermi surface, $\alpha_{CF}$ is related to the intrinsic metric, 
and can be thought of as a coordinate rescaling which yields anisotropic model wavefunctions\cite{Qiu2012,BoYang2012} which can then be used to probe the anisotropy of a gapped state. 
An open question is whether a generalization of such rescaling can be accomplished for $N>2$, i.e. whether model anisotropic wavefunctions with higher-order rotational symmetry can be defined in a unique way, and how they would depend on the anisotropy parameter $\alpha_{CF}$ of the parent composite Fermion state.

\acknowledgments

We acknowledge useful conversations with B. Bradlyn, Z. Papic, F. D. M. Haldane, J. Wang and M. Shayegan. 
We especially thank R. Mong and M. Zaletel for providing the DMRG libraries used in this work. This work was supported by Department of Energy BES Grant DE-SC0002140.

\bibliography{nfold}
\end{document}